\documentclass[twocolumn,aps,pre,superscriptaddress,amsmath,amssymb,amsfonts,nofootinbib]{revtex4-1}

\usepackage{graphicx}% Include figure files
\usepackage{float}
\usepackage{color}
\usepackage{xcolor}
\usepackage{soul}
\usepackage{dcolumn}% Align table columns on decimal point

\graphicspath{{./figures/}}

\begin{document}
\title{
Memory switching due to thermal noise in amorphous solids subject to cyclic shear
}

\author{Debjyoti Majumdar}
\affiliation{Environmental Physics, Jacob Blaustein Institutes for Desert Research, Ben-Gurion University of the Negev, Sede Boqer Campus 84990, Israel}

\author{Ido Regev}
\email[E-mail: ]{regevid@bgu.ac.il}
\affiliation{Environmental Physics, Jacob Blaustein Institutes for Desert Research, Ben-Gurion University of the Negev, Sede Boqer Campus 84990, Israel}
\date{\today}

\begin{abstract}
The discovery that memory of particle configurations and plastic events can be stored in amorphous solids subject to oscillatory shear has spurred research into methods for storing and retrieving information from these materials. However, it is unclear to what extent the ability to store memory is affected by thermal fluctuations and other environmental noises, which are expected to be relevant in realistic situations. Here, we show that while memory has a long lifetime at low temperatures, thermal fluctuations eventually lead to a catastrophic loss of memory, resulting in the erasure of most or all of the stored information within a few forcing cycles. We observe that an escape from the memory-retaining state (limit cycle) is triggered by a change in the switching of plastic events, leading to a cascade of new plastic events that were not present in the original limit cycle. The displacements from the new plastic events change the particle configuration which leads to the loss of memory. We further show that the rate of escaping from a limit cycle increases in a non-Arrhenius manner as a function of temperature, and the probability of staying in a limit cycle decays exponentially with an increase in the shearing frequency. These results have important implications for memory storage since increasing the temperature offers a means of effectively erasing existing memories and allowing for the imprinting of new ones that can then be stored for a long time at low temperatures.
\end{abstract}

\maketitle

\section{Introduction}
The study of mechanical memory in materials is an emerging field \cite{keim2019memory,bense2021complex,pashine2019directed,shohat2022memory,fiocco2014encoding,Keim2021Multiperiodic,merrigan2022emergent,royer2015precisely,khirallah2021yielding,jules2022delicate,liu2021fate,pine2005chaos} that is expected to play an important role in applications ranging from soft robotics \cite{laschi2016soft} to materials with tunable properties \cite{chen2021reprogrammable} and mechanical computation \cite{kwakernaak2023counting}. One important type of material that was shown to exhibit mechanical memory is amorphous solids \cite{regev2013onset,priezjev2013heterogeneous,schreck2013particle,keim2013yielding,kawasaki2016macroscopic}. Amorphous solids subject to low amplitude periodic straining at athermal, quasistatic conditions can be trained to remember past configurations and different training protocols \cite{keim2019memory,adhikari2018memory,fiocco2014encoding,mukherji2019strength}. However, under most conditions and in most environments, amorphous solids are not sheared in athermal quasistatic conditions and are subject to random vibrations of thermal, mechanical, or seismic origin.
It is, therefore, desirable to understand how such vibrations affect the formation of new memory and the longevity of stored memory.
%The dynamics of an amorphous solid is determined by its underlying rugged energy landscape. Memory as described by the cyclic recurrence of the same structural configuration, is a consequence of the fact that the system dynamics is confined to a small region of its phase space which results in self-intersection of the phase space trajectories leading to limit cycles. However, for larger shearing amplitudes the self-intersection fails and the system transcends into a diffusive behavior. In the energy landscape picture, the repetitive steady-state corresponds to a localization of the system around minimas and the system jumps between such minimas during a cycle. The rugged energy landscapes are generally full of such minimas separated by large barriers, therefore, not accessible by athermal systems. However the situation is expected to change as we introduce temperature in the presence of external shearing. Therefore, this study can be regarded as an indirect probe in understanding the associated energy landscape.

In amorphous solids subject to athermal quasistatic shear (AQS) periodic straining has the form of increasing the strain in small straining steps and minimizing the strain after each deformation step. This repeats until a maximal (minimal) strain of $\gamma_{max}$ ($-\gamma_{max}$) is obtained. The strain is then changed in an alternating manner of the form $0\rightarrow\gamma_{max}\rightarrow -\gamma_{max}\rightarrow 0$ which comprises one forcing cycle. If the amplitude $\gamma_{max}$ is smaller than the yielding amplitude, the system eventually reaches a state (usually referred to as a ``limit cycle''), where the particle trajectories repeat after one or more forcing cycles \cite{szulc2022cooperative,Keim2021Multiperiodic,ness2020absorbing,Lindeman2021Multiple,szulc2020forced}. 
Inside such a limit cycle, plastic deformation is mediated by a finite set of hysteretic plastic elements called ``soft spots''  (sometimes referred to as ``shear transformation zones'') \cite{manning2011vibrational,STZ,bouchbinder2007athermal}. Each soft spot switches at a given strain, and some can switch back at a different strain. When a soft spot switches, it creates a strain field that affects the switching fields of other soft spots. When the system reaches a limit cycle, soft spots switch back and forth, and at the end of the limit cycle, they all go back to their zero-strain states \cite{mungan2019networks}. Both particle positions and the switching order of soft spots contain memory that can be retrieved using suitable methods. At athermal conditions, the order of soft spot switching is entirely deterministic, and leaving a limit cycle is impossible. Intuitively, we expect that thermal fluctuations will allow the change of switching ordering and/or the switching of soft spots that do not participate in the limit cycle, which will cause the system to exit the limit cycle. However, it is unclear if such thermally activated switching causes a large or small change in the limit cycle and if such activation involves an Arrhenius-type process or a more complicated thermal activation. 

The current knowledge regarding the formation and retention of mechanical memory at finite temperatures is rather limited as most of the work on amorphous solids under oscillatory shear at finite frequencies and temperatures, focused on post-yield failure fatigue \cite{sha2015cyclic,cochran2022slow,priezjev2020delayed,lindstrom2012structures,bhowmik2022fatigue} and not on the response of memory to such conditions. It is known that amorphous solids subject to oscillatory shear can reach a limit cycle in the presence of small thermal fluctuations \cite{priezjev2016reversible,nagamanasa2014experimental}, but it was found that they exhibit a reduced degree of mechanical reversibility when subjected to a single forcing cycle, compared to the athermal case \cite{farhadi2017shear}. In another example, it was shown that thermal fluctuations can cause systems to leave a limit cycle that was prepared by athermal shearing and to reach a new, stable, limit cycle \cite{regev2013onset}. However, to the best of our knowledge, there are no detailed numerical studies of the behavior of limit cycles under a range of temperatures and frequencies and only one theoretical study where it was assumed that such systems leave a cycle due to standard Arrhenius activation \cite{mungan2021metastability}. 

In this manuscript, we present quantitative results from non-equilibrium molecular dynamics (NEMD) simulations of shearing of amorphous solids at different temperatures and frequencies that clearly show that limit cycles containing mechanical memory retain configurational memory to a very good degree even at finite temperatures and strain rates. However, we find that after a certain time, thermal fluctuations induce a change in the switching of soft spots, which causes new soft spots to appear that in turn leads to a rapid erasure of memory. We then show that the escape rate as a function of temperature exhibits a sub-Arrhenius behavior and that the probability of leaving a limit cycle as a function of frequency exhibits an exponential decay. We further show that the time to leave a limit cycle exhibits a power-law distribution. We then discuss the implications for understanding the physics of plastic deformation in amorphous solids and storing memory in such systems.

\section{Memory in the presence of thermal noise}
%
%\dm{In this section, we qualitatively discuss the immediate effects of thermal fluctuations on the memory retaining capability of our amorphous solid.} 
To study the effect of different temperatures and frequencies on memory stability in amorphous solids subject to oscillatory shear, we simulated two-dimensional binary mixtures of particles interacting via a Lenard-Jones potential. Each system comprises $N$ particles enclosed in a square simulation box of size $L$ with periodic boundary conditions. We used two different number densities: $\rho=0.8$ with linear size $L=32$, and $\rho=0.7$ with $L=35$.
Each realization was prepared by quenching a liquid from high temperature to $T=10^{-4}$ and applying oscillatory strain of the form:
\begin{equation}
\gamma = A\sin(\omega t)\,,
\end{equation}
using the LAMMPS \cite{plimpton1995fast} implementation of the SLLOD algorithm \cite{evans1984nonlinear} together with Lees-Edwards periodic boundary conditions \cite{lees1972computer}. Here $\gamma$ is the applied strain, $t$ is time, $A$ is the strain amplitude and $\omega$ the shearing frequency.

\begin{figure}[t]
\centering
\includegraphics[width=\linewidth]{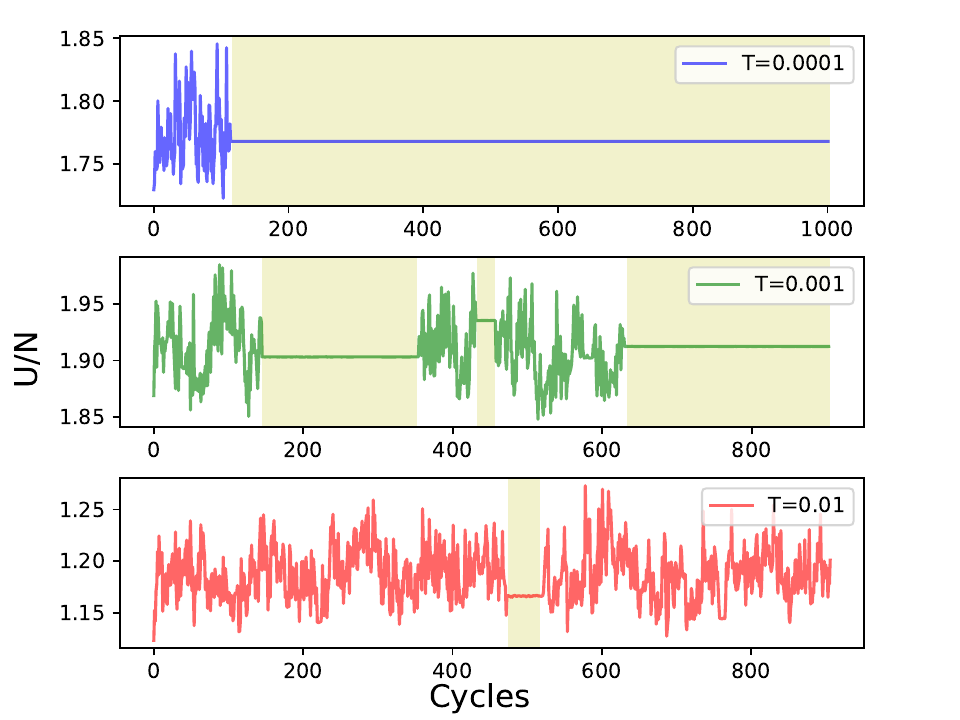}
\caption{Zero-strain (stroboscopic) energy per particle in amorphous solids subjected to oscillatory shear at a frequency $\omega=0.001$ and amplitude $A=2.2$, for three different temperatures $T=10^{-4}$, $10^{-3}$ and $10^{-2}$ in a system of $N=841$ particles at a density of $\rho=0.7$. Periodic states (highlighted) are manifested by a constant zero-strain potential energy. Of importance is the fact that at intermediate temperatures the system is able to jump in and out of the limit cycle for multiple times as shown for $T=0.001$.}
\label{fig1}
\end{figure}

In Fig.~\ref{fig1}, we show examples of the zero-strain (stroboscopic) potential energy per particle ($U/N$) of an amorphous solid sheared at three different temperatures, each differing from the other by an order of magnitude, though still well below the liquefaction point. %and all of which are significantly below the glass transition temperature ($T_g=0.31$). 
In the top-most sub-figure, the temperature is the lowest ($T=10^{-4}$ in simulation units) and the system reaches a periodic state (highlighted background) after a transient of about $100$ training cycles.
%%%%%%%%%%%%%%%%%%%%%%%%%%%%%%%%%%%%%%%%%%
%%%%%%%%%%%%%%%%%%%%%%%%%%%%%%%%%%%%%%%%%%
For this temperature, the limit cycle reached is stable and the solid stays in the same state, indicated by the fact that the potential energy remains unchanged after each cycle for an extended period of time. %, which therefore appears as a straight line. 
This observation suggests that the memory obtained is stable against the thermal fluctuations at that temperature and strain rate. In the second case (middle sub-fig of Fig.~\ref{fig1}), where the temperature is higher  ($T=10^{-3}$), the system also reaches a periodic state after about 150 forcing cycles and stays in the same limit cycle for an extended period of time. However, in this case, after about $200$ more forcing cycles, the system leaves the periodic state, starts a new random transient, and after $50$ more forcing cycles, reaches a new limit cycle. Nevertheless, the new limit cycle is also unstable, and the system leaves it after a relatively small number of cycles. The dynamics involving jumping in and out of limit cycles keep repeating, seemingly forever. 
%\dm{Note that these intermittent limit cycles are degenerate, they are configurationally different but contains memory of the same amplitude}. %IR memory storage is not only of the amplitude
In this case, it is easy to observe that the memory has a limited, random expiration time. Lastly, at an even higher temperature ($T=10^{-2}$), shown in the bottom sub-fig of Fig.~\ref{fig1}, the system spends most of the time in non-repetitive dynamics but still manages to find a short-lived limit cycle. A question arises as to the extent to which the particle configuration changes when jumping between limit cycles, and this will be discussed below.
%%%%%%%%%%%%%%%%%%%%%%%%%%%%%%%%%%%%%%%%%%
\begin{figure}[t]
\centering
\includegraphics[width=\linewidth]{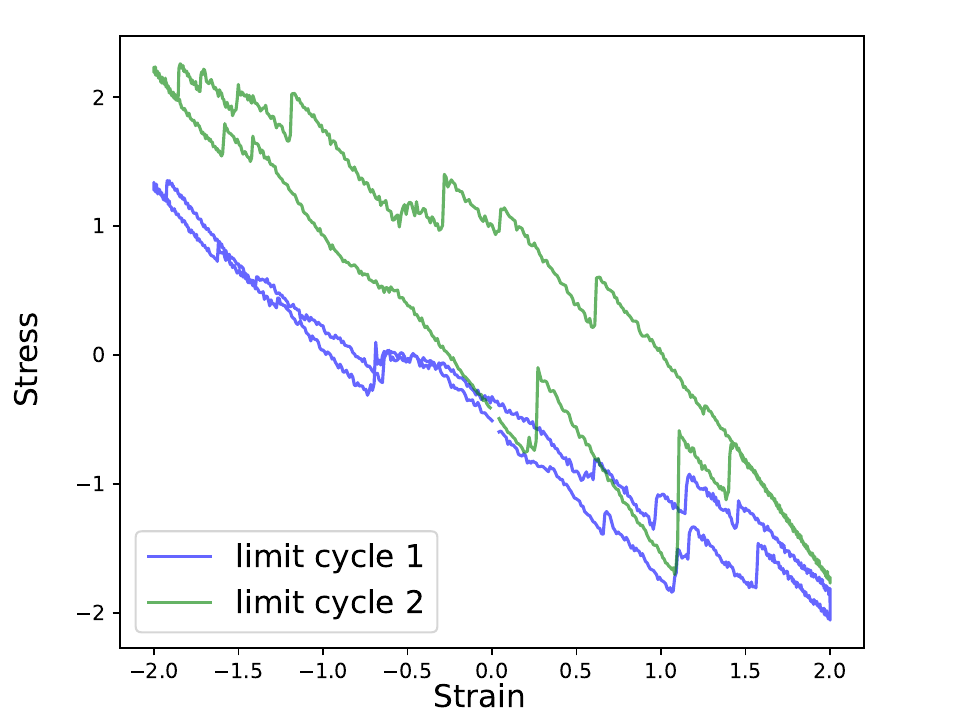}
\caption{
The hysteresis curves of two limit cycles that were reached during the same simulation at $T=0.003$. Note that the transient connecting them is not shown.
%The different limit cycles are marked with different colors. 
%(b) Stress vs steps showing jumping out of limit cycle at $T=0.007$. In both cases the amplitude was $A=2$ the system size was $N=841$ particles and the density was $\rho=0.7$. \rv{IR: please change x-axis to time}
}
\label{fig2}
\end{figure}
%%%%%%%%%%%%%%%%%%%%%%%%%%%%%%%%%%%%%%%%%%
\section{Memory loss due to thermal noise}
In Fig.~\ref{fig2}, we show two limit cycles connected by a random transient (not shown). The plastic events and thermal noise are visible in both cycles, and it seems that most plastic events in the two cycles are different. However, one may wonder to what extent the particle configurations reached after the system escapes from a limit cycle differ from those inside the limit cycle. To check this, we calculated the self-part of the intermediate scattering function, given by:
\begin{equation}
\mathcal{F}_s({\bf k},t)=\frac{1}{N}\sum_{i=1}^N \cos({\bf k}\cdot[{\bf r}_i(t)-{\bf r}_i(0)]),
\end{equation}
where ${\bf r}_i$ is the coordinate of the $i$th particle and ${\bf k}$ is a wave-vector. When ${\bf k}$ is chosen to have a size comparable to the average distance between particles, the decay of the function indicates diffusion away from the first shell of neighboring particles (this function was previously used to study self-diffusion close to the glass transition) \cite{donati1998stringlike,berthier2011theoretical}. 
%
%%%%%%%%%%%%%%%%%%%%%%%%%%%%%%%%%%%%%%%%%%
\begin{figure}[t]
\centering
\includegraphics[width=\linewidth]{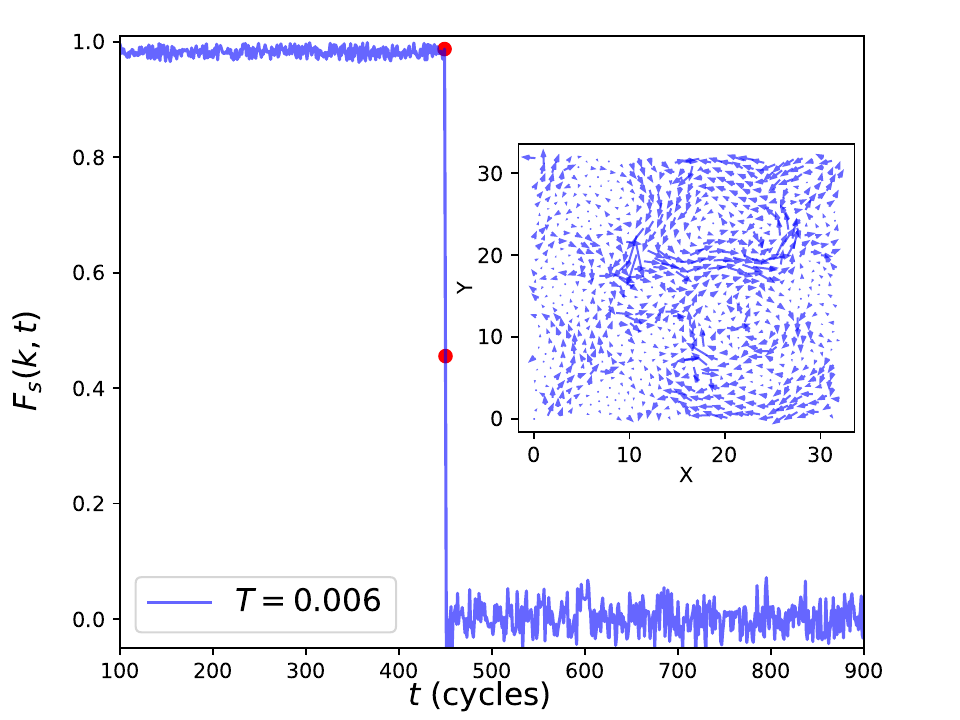}
\caption{Self-part of the intermediate scattering function for a single sample at $T=0.006$ exhibiting a step-like change from $1$ inside a limit cycle to $0$ when the system escapes the limit cycle. (Inset) The corresponding displacement between the zero-strained configurations across the point where $F_s(k,t)$ decays to zero.}
\label{fig3}
\end{figure} 
%%%%%%%%%%%%%%%%%%%%%%%%%%%%%%%%%%%%%%%%%%
%%%%%%%%%%%%%%%%%%%%%%%%%%%%%%%%%%%%%%%%%%
\begin{figure}[b]
\includegraphics[width=.7\linewidth]{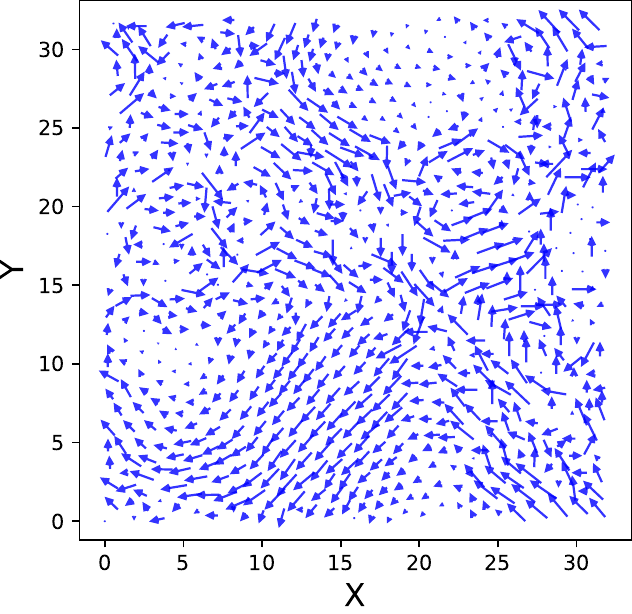}
\caption{The displacements of particles between zero-strain configurations before and after a thermal escape from a limit cycle. }
\label{fig4}
\end{figure} 
%%%%%%%%%%%%%%%%%%%%%%%%%%%%%%%%%%%%%%%%%%
In Fig~\ref{fig3}, we show a typical case, where we use $k_x=k_y=2\pi/\sigma_{\alpha}$ with $\sigma_A=1$ for the small particles and $\sigma_B=1.4$ for the large particles. Intriguingly, $\mathcal{F}_s(k,t)$ changes sharply from $1$ to $0$ as the system jumps out of the limit cycle (Fig.~\ref{fig3}) in a manner resembling a Heaviside step function. This indicates an abrupt change in the configuration, which implies that the memory stored in the limit cycle is lost in a small number of forcing cycles following the point from which the system leaves the limit cycle. To understand the origin of this behavior, we looked at the displacements between the zero strain configurations before and after an escape from a limit cycle (see Fig.~\ref{fig3} inset and \ref{fig4}). We observed that the displacements due to the new plastic events cover a significant part of the material, which explains the abrupt change in $\mathcal{F}_s(k,t)$. 

\begin{figure}[t]
\includegraphics[width=.8\linewidth]{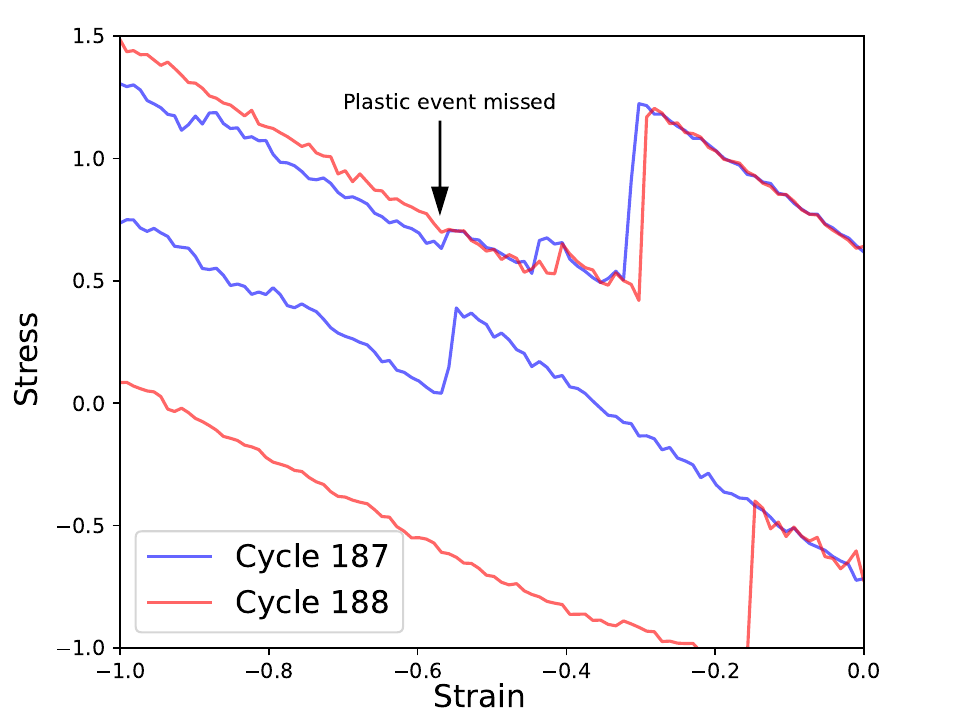}
\caption{Hysteresis curves of two consecutive cycles where the system leaves the limit cycle during cycle 188. In both cycles, the temperature was $T=7\times 10^{-4}$ and the amplitude was $A=2$ and the frequency was $\omega=0.001$. The transition out of the limit cycle occurs due to one plastic event (marked with an arrow) being ``missed'' which causes plastic events that were not present in cycle 187 to occur.
}
\label{fig5}
\end{figure}

To identify the point of departure from the limit cycle, we plot the hysteresis curve for two consecutive cycles, one inside and another immediately outside the limit cycle in Fig~\ref{fig5}. During forcing cycle 187 (counting from the onset of shearing), the system is still inside a limit cycle, but at forcing cycle 188 leaves the limit cycle. We can see that the deviation of cycle 188 from the stress-strain curve of 187 starts when one plastic event (marked by an arrow) does not occur. This, next, triggers the switching of new plastic events that were not present in 187 and prevents plastic events that were present in 187 from appearing. Looking at the displacement fields of the individual plastic events (not shown), we 
found that the displacements due to these new events accumulate during the cycle such that they cover a large part of the simulation box. We, therefore, conclude that the sharp structural change observed in Fig~\ref{fig3} is due to the combination of two effects - a plastic event that was not switched due to thermal fluctuations allows the switching of a new plastic event that was not switched as part of the limit cycle. This leads other plastic events that did not take part in the cycle to occur, which causes the particle configuration to change such that the memory is erased.
%
%%%%%%%%%%%%%%%%%%%%%%%%%%%%%%%%%%%%%%%%%%
%%%%%%%%%%%%%%%%%%%%%%%%%%%%%%%%%%%%%%%%%%

\begin{figure}[h]
\includegraphics[width=.8\linewidth]{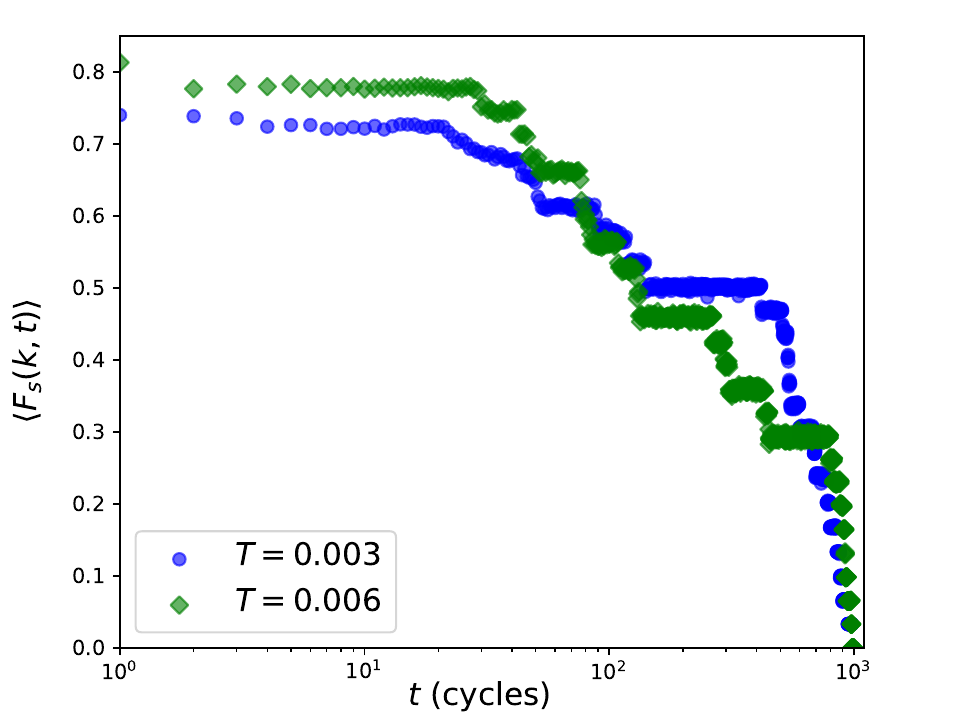}
\caption{Self part of the intermediate scattering function averaged over $30$ samples for two different temperatures $T=0.003$ and $0.006$. Here $t$ is the number of cycles.
}
\label{fig6}
\end{figure}

\begin{figure}[b]
\includegraphics[width=\linewidth]{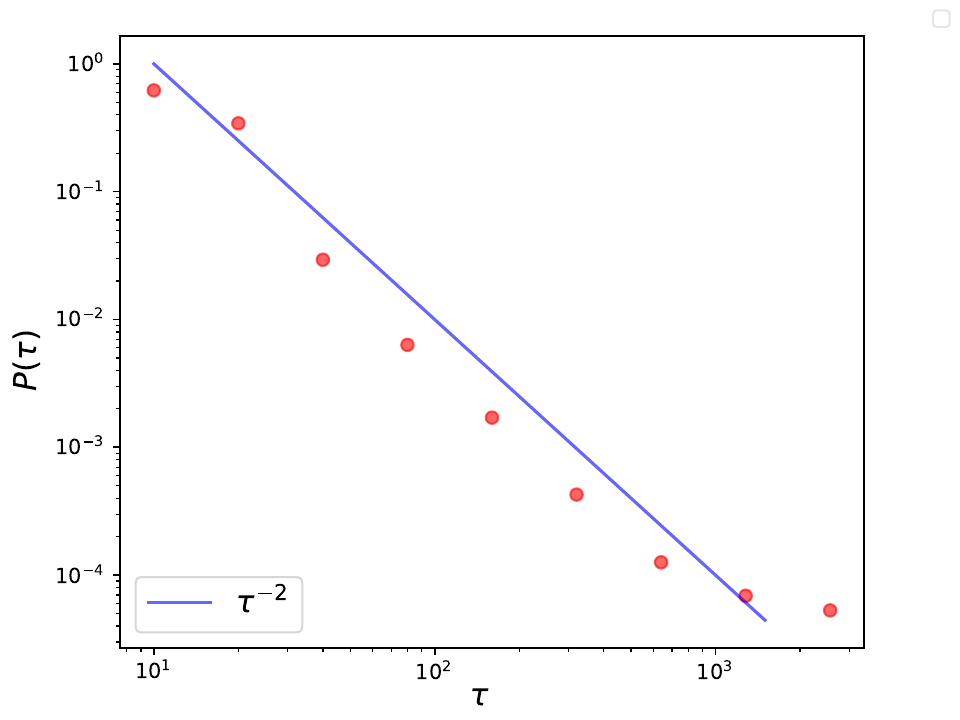}
\caption{The distribution of limit cycle longevity $\tau$ exhibiting a power-law decay with exponent $2$. 
%\dm{Absolute frequency of each bin is first divided by the corresponding bin width and then normalized by the total sum.}
}
\label{fig7}
\end{figure}

In Fig~\ref{fig6} we show $\mathcal{F}_s(k,t)$ averaged over 30 realizations for two different temperatures $T=0.003$ and $0.006$. The configuration corresponding to the first limit cycle for each individual sample is taken as the reference configuration for that sample i.e. ${\bf r}(0)$. In all of the realizations, we observed the same step-like switching observed in Fig~\ref{fig3} where $\mathcal{F}_s(k,t)$ switches abruptly from $1$ to $0$ at the vicinity of a specific forcing cycle. However, in different realizations, the switching occurs at different times, which causes the more gradual decay observed in Fig~\ref{fig6}. 
In fact, for the temperatures shown, the distribution of switching times was found to be a power-law, as shown in Fig~\ref{fig7}. We verified that when the number of samples is small, summing a set of step-functions (realizations) that switch at a random time (cycle number) reproduces behavior similar to the one shown in Fig~\ref{fig6}, while when the number of step-functions increases, $\mathcal{F}_s(k,t)$ exhibits a power-law decay, as expected. From  Fig~\ref{fig6} it seems that the temperature has an effect on the average decay rate. A quantitative analysis of this difference is discussed in the next section.

%%%%%%%%%%%%%%%%%%%%%%%%%%%%%%%%%%%%%%%%%%
%%%%%%%%%%%%%%%%%%%%%%%%%%%%%%%%%%%%%%%%%%

%%%%%%%%%%%%%%%%%%%%%%%%%%%%%%%%%%%%%%%%%%
%%%%%%%%%%%%%%%%%%%%%%%%%%%%%%%%%%%%%%%%%%
%
\section{Memory stability at different temperatures}
\begin{figure}[t]
\includegraphics[width=\linewidth]{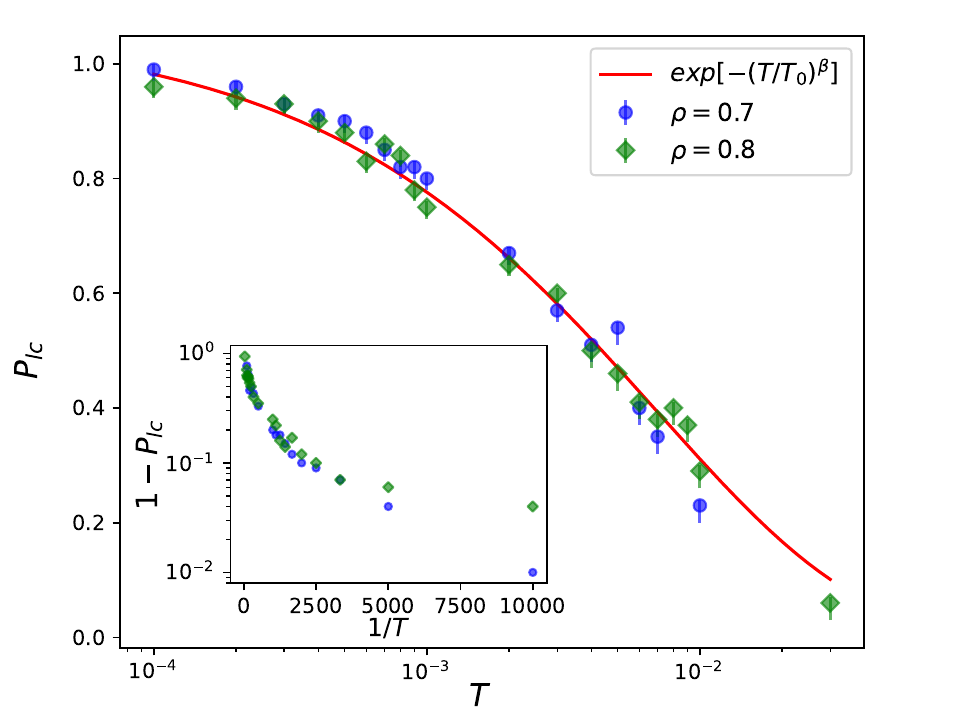}
\caption{The probability $P_{lc}$ to stay in the initial limit cycle as a function of temperature $T$ of a system consisting of $N=841$ particles at a density of $\rho = 0.8$, and $N=900$ particles at a density of $\rho=0.7$. The initial samples were prepared at $T_i=10^{-4}$, $\omega_i=0.001$, and amplitude $A=2.1$. A fit (continuous red line) indicates a decaying exponent behavior with an exponent of $\beta = 0.59$ and $T_0=0.007$.
Inset: Arrhenius plot of the escape rate $1 - P_{lc}$ as a function of $1/T$ showing a deviation from Arrhenius behavior at low temperatures. 
} 
\label{fig8}
\end{figure}

\begin{figure}[t]
\includegraphics[width=\linewidth]{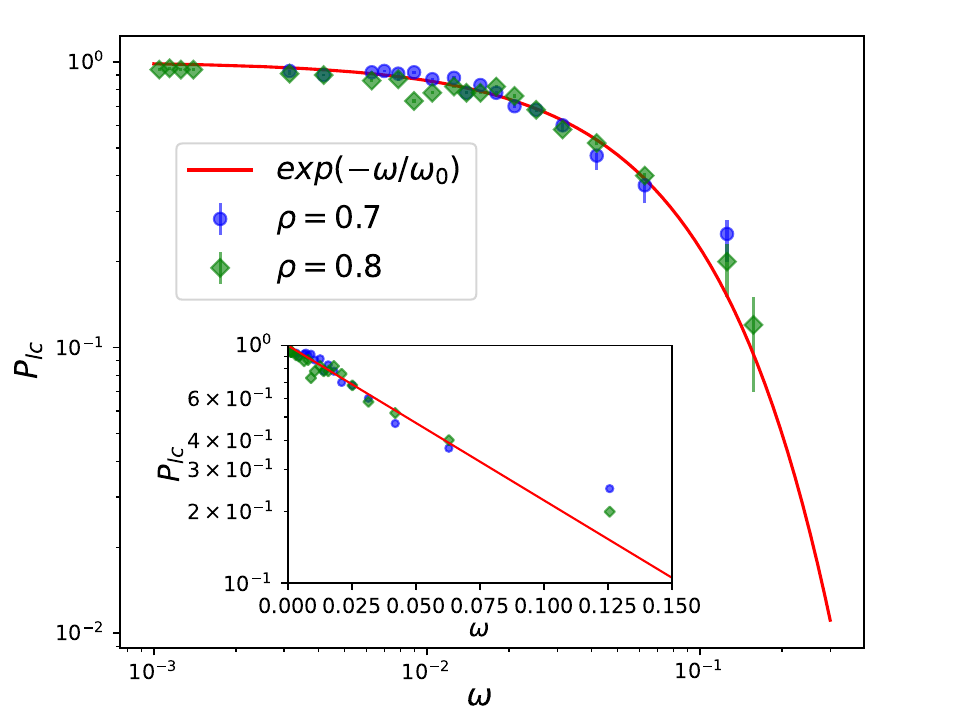}
\caption{$P_{lc}$ as a function of frequency ($\omega$) exhibiting an exponential decay. The initial configurations are prepared at $T_i=10^{-4}$ and $\omega_i=0.001$, amplitude $A=2.1$. (Inset) Same as the outer plot but in the semi-log axis. The continuous line is a fit with $\omega_0=0.067$.
} 
\label{fig9}
\end{figure}

As demonstrated in Fig.~\ref{fig1}, temperature strongly affects the dynamics, which is reflected in the limit cycle escape times. To quantify the effect of temperature on the escape time, we measured the average escape rate using 100 realizations. The initial realizations were prepared by applying 1000 forcing cycles at an amplitude $A=2.1$, frequency $\omega_i=10^{-3}$ and temperature $T_i=10^{-4}$. We verified that all the realizations reached a limit cycle which was used as the initial state. The prepared samples were next subjected to shearing with the same amplitude and frequency but using a temperature $T\geq T_i$. We then recorded the fraction of realizations that stayed in the initial limit cycle after 10 forcing cycles $P_{lc}$ as a function of temperature. $P_{lc}(T)$ is shown in Fig.~\ref{fig8} for the two densities, $\rho=0.8$, and $\rho=0.7$. We found that $P_{lc}$ exhibits a stretched exponential decay $P_{lc}\sim\exp(-(T/T_0)^\beta)$ with an exponent $\beta=0.59$ and a temperature scale $T_0=0.007$. In the inset, we show an Arrhenius plot of the escape rate $1 - P_{lc}$ as a function of $1/T$. We can see that at low temperatures, the escape rate is Arrhenius-like, but at temperatures above $T \sim 10^{-3}$, we see a strong deviation from Arrhenius behavior for both densities.
%
%To understand the change in behavior we calculate the P\'eclet number, that in LJ units is:
%\begin{equation}
%Pe = \frac{A u}{D} = \frac{A^2\omega}{T/\gamma} = \gamma\frac{A^2\omega}{T}
%\end{equation}
In Fig.~\ref{fig9}, we show the same quantity but at a constant temperature $T_i=10^{-4}$ and a varying frequency $\omega$. In this case, $P_{lc}(\omega)$ exhibits an exponential decay. This indicates that the stability of limit cycles is very sensitive to temperature increases, while it is somewhat less sensitive to an increase in the shearing frequency. 
%This could have implications for memory storage \rv{IR: need to think of something}.

%is changed but the probability to continue in the previous steady-state decays as an exponential function of the $\omega$. From  here we can define a characteristic frequency where the possibility of memory retention is $e^{-1}$ times of the low temperature which is about $T=0.006(9)$.

%
%%%%%%%%%%%%%%%%%%%%%%%%%%%%%%%%%%%%%%%%%%
%%%%%%%%%%%%%%%%%%%%%%%%%%%%%%%%%%%%%%%%%%
%%%%%%%%%%%%%%%%%%%%%%%%%%%%%%%%%%%%%%%%%%
%%%%%%%%%%%%%%%%%%%%%%%%%%%%%%%%%%%%%%%%%%

\section{Discussion}
In this study, we investigated the impact of finite temperatures and strain rates on the formation and stability of memory in amorphous solids subject to oscillatory shear. Our findings indicate that memory of particle positions and plastic events can be stored accurately for a large range of temperatures and frequencies and that the main difference between high and low temperatures is the amount of time for which memory is stored rather than the quality of the memory stored.
Our simulations indicate that when the system is inside a limit cycle, the main effect of thermal fluctuations is to shift the strain values at which different plastic events are switched during the cycle. However, after some time, these fluctuations become large enough to cause a ``scrambling'' of the switching fields of the soft spots, an effect that was attributed to soft spot interactions in previous work \cite{bense2021complex,szulc2022cooperative}. This then causes the original soft spots to become disabled and a completely new set of soft spots to switch. It is not clear how the switching of one new soft spot can cause such an extensive loss of memory but it is most likely a result of the elastic interactions between soft spots, interactions that were shown to cause the shifting of the switching strains \cite{mungan2019networks}. Another intriguing result is the sub-Arrhenius decay of the rate at which the system escapes a limit cycle. Since the set of energy minima visited by the system during a limit cycle cannot change, as long as the system is inside the same limit cycle, this cannot be a result of a change in the energy landscape region visited (contrary to what is found in supercooled liquids \cite{debenedetti2001supercooled,sastry1998signatures}) but is most likely a result of the combined effects of inertia and thermal fluctuations. 
We believe that most of the computer experiments that we performed should be accessible to actual experiments in colloidal glasses and it will be interesting to see how experiments and simulations compare.
Lastly, we believe that the abrupt memory loss may also be useful as a means of erasing an existing memory and preparing the system for the imprinting of a new memory by ramping the temperature up and down. Future work will focus on understanding the relationship between temperature, frequency, and amplitude on memory lifetime, the origin of the power-law behavior, $T_0$ and $\omega_0$, and the mechanism behind the abrupt memory loss.\\

{\it Acknowledgements}:
DM and IR were supported by the Israel Science Foundation (ISF) through Grant No. 1301/17. DM and IR would like to thank Itamar Procaccia for useful comments.

\section{Materials and methods} \label{sc2}
\subsection{Model and simulation details} 
We simulated a 50:50 binary mixture of two-dimensional particles interacting via a Lenard-Jones (LJ) pair-wise potential using LAMMPS \cite{plimpton1995fast}:
\begin{equation}
V_{\alpha \beta}(r) = 4\epsilon_{\alpha \beta}\left[\left(\frac{\sigma_{\alpha \beta}}{r} \right)^{12} - \left(\frac{\sigma_{\alpha \beta}}{r} \right)^{6} \right]~~~(r<r_c),
\end{equation}
where $\alpha,\beta\in\{A,B\}$,  $\epsilon_{AA}=\epsilon_{AB}=\epsilon_{BB}=1$, $\sigma_{AA}=1$, $\sigma_{AB}=1.2$, and $\sigma_{BB}=1.4$, whereas the masses of the two particle sizes are equal $m_A=m_B=1$. The cutoff radius is taken to be $r_{c,\alpha\beta}=2.5 \sigma_{\alpha \beta}$. 
In all simulations we predefined the linear system size $L$ and the number density $\rho$ whereas the number of particles $N$ was determined from the relation $\rho=N/L^2$. We used two densities, $\rho=0.8$ with $L=32$, and $\rho=0.7$ with $L=35$ which keeps the particle number roughly equal. Note that the density is well above the jamming density $\rho_J=0.64$ in LJ systems. We choose half of the particles to be $1.4$ times larger than the other half which inhibits crystallization \cite{perera1999stability}. 
All simulation results are reported in LJ units, with the time - being measured in units of $\tau=\sqrt{m\sigma^2/\epsilon}$.

Each amorphous solid realization is first equilibrated for $10^6$ time steps of value $\delta t=10^{-4}$ at a constant volume $V$ and a high temperature $T=2$. Next, the temperature is ramped down to the target temperature of $T=10^{-4}$ with - additional equilibration steps at two intermediate temperatures, $T=1$ and  $T=0.1$ for $10^6$ steps each. Finally, the system is again equilibrated for another $10^6$ time steps at the target temperature. In order to simulate time-dependent shearing, we used the Lees-Edwards boundary conditions \cite{lees1972computer} coupled with the SLLOD equations of motion \cite{evans1984nonlinear} where the temperature was maintained using a Nos\'e-Hover thermostat \cite{nose1984unified,allen1989computer} . Time-periodic shear is imposed along the x direction as $\gamma(t)=A\sin(\omega t)$, where $A$ is the strain amplitude and $\omega$ is the frequency, using the same time step as $\delta t=10^{-4}$. 

\subsection{Limit cycle detection}

To detect the onset of a limit cycle, we looked at the potential energy per particle ($U/N$) of the zero-strain configuration after each shearing cycle, and calculated the variance of the zero-strain configurational energy for three consecutive cycles, using the expression:
\begin{equation}
\mbox{Var}(u) =\frac{1}{3}\left[(u_n-\mu)^2+ (u_{n-1}-\mu)^2 + (u_{n-2}-\mu)^2\right],
\end{equation}
where $\mu=\frac{1}{3}(u_n+u_{n-1}+u_{n-2})$ is the mean zero-strain energy of the last three cycles $n$, $n-1$ and $n-2$. At low temperatures, the variance inside a limit cycle could be as low as Var$(u)\sim 10^{-11}$ while outside a limit cycle it is Var$(u)\sim 10^{-5}$. 

%Besides using the zero-strain energy variance of the last few cycles, we can also look at the mean-squared displacement (MSD) with reference to the starting configuration according to Eq. \ref{eq3}, where the MSD increases linearly with the number of shearing cycles in the transients and eventually reaches a plateau in the SSLC [Fig. \ref{fig1}]. Other than that, we can also look at the overlap between consecutive zero-strain configurations to detect the limit cycles. These two methods, should be more suitable for experimental purposes, since in experiments the configurations are more readily available than thermodynamic variables such as $U$. 
%
\bibliography{NSF}
\end{document}